\documentclass[lettersize,journal]{IEEEtran}
\usepackage{arydshln}
\usepackage{enumitem}
\usepackage{amssymb}
\usepackage{amsmath,amsthm}
\usepackage{algorithm}
\usepackage{algorithmic}
\usepackage{array}
\usepackage[caption=false,font=normalsize,labelfont=sf,textfont=sf]{subfig}
\usepackage{textcomp}
\usepackage{stfloats}
\usepackage{url}
\usepackage{verbatim}
\usepackage{graphicx}
\usepackage{booktabs}
\usepackage{rotating}
\usepackage{balance}
\usepackage{tabularx}
\usepackage{bm}
\newcolumntype{Y}{>{\centering\arraybackslash}X}
\usepackage{tikz}
\usetikzlibrary{decorations.pathreplacing,calc}
\newcommand{\tikzmark}[1]{\tikz[overlay,remember picture] \node (#1) {};}

\newcommand*{\AddNote}[4]{%
    \begin{tikzpicture}[overlay, remember picture]
        \draw [decoration={brace,amplitude=0.7em},decorate,thick,black]
            ($(#3)!(#1.north)!($(#3)-(0,1)$)$) --  
            ($(#3)!(#2.south)!($(#3)-(0,1)$)$)
                node [align=center, text width=2.5cm, pos=0.5, anchor=west] {#4};
    \end{tikzpicture}
}%

\newtheorem{thm}{Theorem}[section]
\newtheorem{lem}[thm]{Lemma}

\theoremstyle{definition}
\newtheorem{defn}{Definition}[section]
\newtheorem{exmp}{Example}[section]
\newtheorem{rem}{Remark}
\begin{document}

\title{Local Distance Query with Differential Privacy}
\author{
        Weihong~Sheng,
        Jiajun~Chen,
        Bin~Cai,
        Chunqiang~Hu,
        Meng Han,
        \and Jiguo Yu
}

\maketitle

\begin{abstract}
Differential Privacy (DP) is commonly employed to safeguard graph analysis or publishing. Distance, a critical factor in graph analysis, is typically handled using curator DP, where a trusted curator holds the complete neighbor lists of all vertices and answers queries privately. However, in many real-world scenarios, such a curator may not be present, posing a significant challenge for implementing differentially private distance queries under Local Differential Privacy (LDP).
This paper proposes two approaches to address this challenge. The first approach generates a synthetic graph by randomizing responses and applies bitwise operations to reduce noise interference. However, like other synthetic graph methods, this approach suffers from low utility. To overcome this limitation, we propose a second approach, the first LDP method specifically designed for distance queries, which captures the global graph structure by continuously aggregating local distance vectors from neighboring vertices. This process enables the accurate updating of global distances. We demonstrate the effectiveness of our method through comprehensive theoretical analysis and experimental evaluations on real-world datasets.
\end{abstract}

\begin{IEEEkeywords}
local differential privacy, distance, graph, neighbor aggregation.
\end{IEEEkeywords}

\maketitle

\section{Introduction}

Amid frequent privacy breaches, ensuring privacy in data collection and analysis has become crucial to addressing public concerns. Without a privacy promise, users may hesitate to provide personal data that could improve service quality. Differential Privacy (DP), introduced formally by Dwork et al. \cite{dwork2006calibrating}, has attracted significant research interest due to its strong privacy guarantees and versatile applicability. Notable applications of DP include deployments by major corporations such as Microsoft \cite{ding2017collecting}, Apple \cite{tang2017privacy} and Google \cite{erlingsson2014rappor}, enhancing the privacy of their services.

Privacy-preserving analysis on graphs has become an important research trend, driven by the widespread use of graph-structured data and growing concerns about data privacy\cite{nguyen2023faster, brito2023global}. However, when it comes to distance queries, such as computing the shortest path between two vertices or all-pairs shortest paths in unweighted graphs, most existing studies rely on the curator model of DP, in which a trusted curator possesses the entire graph and answers queries in a privacy-preserving manner \cite{sealfon2016shortest, fan2022distances, fan2022private, chen2023differentially}. The existence of such a curator is not guaranteed in practical scenarios, and even if one is present, data owners may incur a trust cost, fearing that the curator could compromise their privacy. Alternatively, Local Differential Privacy (LDP) eliminates the need for a trusted curator by sanitizing data before it leaves the user's control. 

Consider a distributed social network where vertices represent users and edges denote sensitive friendships, with each user holding their own friend list. A service provider maintains the network and often needs to collect statistics, such as average distance, to improve its services. While users may be willing to contribute, they do so only if privacy is preserved. Furthermore, consider the two cases:
\begin{itemize}
    \item \textbf{Case $1$: Analyzing contact networks during an epidemic outbreak.} Distance queries on contact networks can help identify $3$-hop contacts or assess transmission risks. However, due to privacy concerns, individuals may be unwilling to share data or may provide inaccurate information.

    \item \textbf{Case $2$: Analyzing communication networks among sensitive populations.} For groups such as investigative journalists or sex workers, communication often occurs via decentralized encrypted networks. The challenge lies in assessing network robustness against censorship or attacks without revealing specific interpersonal connections.
\end{itemize}

Therefore, a significant challenge remains: how to perform differentially private distance queries without relying on a curator who possesses all neighbor lists.

One possible solution to this challenge is to generate synthetic graphs. Specifically, this involves privately collecting degree and other auxiliary information from each local vertex to construct a synthetic graph using random graph generation techniques \cite{aiello2000random,seshadhri2012community}. However, this approach tends to introduce substantial noise, which can severely distort the community structure \cite{qin2017generating}. As an alternative to synthetic graph approaches, one strategy is to directly collect statistical information from each local vertex. This approach avoids the complexity of synthetic graphs but faces challenges in accurately capturing distances, which require knowledge of the global graph structure, unlike simple metrics such as subgraph counts.

In this paper, we address the challenge of answering distance queries under LDP and propose two innovative approaches. The first approach involves generating synthetic graphs through randomized responses, with the aim of preserving the global graph structure. To mitigate the damage to the graph structure caused by randomized responses, we employ post-processing techniques using $\textbf{AND}$ and $\textbf{OR}$ operations. However, this approach still suffers from inherently low utility, as the added noise significantly disrupts structural integrity.

To address this limitation, we propose a second approach inspired by GNN. In this approach, each vertex maintains a local distance vector, initially containing distances only to its immediate neighbors. By continuously aggregating these local distance vectors from their neighbors, the vertices can reconstruct the global structure and obtain the global distances.

The main contributions of this paper are summarized as follows:
 \begin{itemize}
    \item We enhance randomized response-based synthetic graph generation using $\textbf{AND}$ and $\textbf{OR}$ operations, which help reduce noise and more accurately preserve the underlying graph density. Furthermore, we provide a detailed theoretical analysis of the density errors introduced by the perturbation process.
    
    \item We design a novel distance aggregation mechanism that reconstructs global distances by iteratively aggregating local distance vectors from neighboring vertices. We further provide theoretical guarantees for both Laplace noise and randomized response, offering guidance on selecting optimal perturbation mechanisms.
    
    \item We demonstrate the effectiveness of the proposed distance query mechanisms through comprehensive numerical evaluations and empirical analysis on real-world datasets.
 \end{itemize}
The rest of the paper is as follows. In Section \ref{sec:Pre}, we show the background knowledge covered in this paper. In Section \ref{sec:gam}, we introduce our synthetic graph generation approach. In Section \ref{sec:nam}, we detail our distance aggregation approach. In Section \ref{sec:exp}, we describe the experimental design and results. In Section \ref{sec:RW}, we present related work. Finally, in Section \ref{sec:con}, we summarize this paper.

\section{Preliminaries}\label{sec:Pre}
\subsection{Problem Definition}

Let $\mathbf{N}_i$ represent the neighbor list of vertex $i \in[n]$, corresponding to the $i$-th row of the adjacency matrix. The query function $f$, a distance query function, returns all-pair distances when applied to $(\mathbf{N}_1, \mathbf{N}_2, \ldots, \mathbf{N}_n)$. Each edge, connecting any vertices $i$ and $j$, is undirected. Given that edges are considered private, vertices avoid disclosing their neighbors when responding to the query $f$. Let $\mathcal{M}$ denote a randomized mechanism that provides answers to all-pair distance queries while preserving privacy. Designing $\mathcal{M}$, which can answer all-pair distances with low error while providing differential guarantee for edges, is the main purpose of this paper.

\subsection{Differential Privacy}
Differential privacy \cite{dwork2006calibrating} is a formal privacy standard that ensures privacy by minimizing the influence of any individual's record on the outcome of a query. Any adversary cannot obtain additional knowledge by observing the query outputs from two databases: one that includes and one that excludes the record of a target individual.

Let us consider the form of DP on the graph. Suppose $G$ is a simple graph with $n$ vertices and $m$ edges. The privacy-sensitive information we aim to protect is the presence of any edge (or relationship) within $G$ during the execution of distance queries on $G$.

\begin{defn}
(Neighboring Datasets\cite{dwork2006calibrating}) Let $G, G^\prime \in \mathcal{G}$ be two undirected graphs. We say $G=(V,E)$ and $G^\prime=(V,E^\prime)$ are neighboring datasets, denoted as $G\sim G^\prime$, if they have the same vertex sets but edge sets differ in one edge: $|E-E^\prime|=1$.
\end{defn}

\begin{defn}
(Edge Differential Privacy\cite{hay2009accurate}) Let $\mathcal{M}: \mathcal{G}\rightarrow \mathcal{S}^n$ be a randomized mechanism. Given any neighboring databases $G$ and $G^\prime$, which differ in one edge, then $\mathcal{M}$ satisfies $\varepsilon$-DP if, for all $S \in \operatorname{Range}(\mathcal{M})$, it holds that
\begin{equation}
    \operatorname{Pr}\left[\mathcal{M}\left(G\right) \in S\right] \leq \exp (\varepsilon) \times \operatorname{Pr}\left[\mathcal{M}\left(G^{\prime}\right) \in S\right].
\end{equation}
\end{defn}
The parameters $\varepsilon$ quantify the privacy loss of $\mathcal{M}$, with smaller value indicating stronger privacy guarantees. Specifically, $\varepsilon$ constrains the probability ratio between the outputs of $\mathcal{M}$ on database $G$ and $G^\prime$. DP can be achieved by adding Laplace noise, if the query $f$ is numeric. 

\begin{defn}
(Laplace Mechanism\cite{dwork2006calibrating}) Let $f:\mathcal{G} \rightarrow \mathcal{S}^n$ be a distance query function. $\mathcal{M}: \mathcal{G} \rightarrow \mathcal{S}^n$ satisfies $\varepsilon$-DP, if
\begin{equation}
    \mathcal{M}(G)=f(G)+\frac{\Delta f}{\varepsilon} X^n,
\end{equation}
where $X^n$ represents $n$ independent and identically distributed (i.i.d.) Laplace random variables and $\Delta f$ is the sensitivity of $f$, which is
\begin{equation}
    \Delta f= \max_{G\sim G^\prime}{ \left\|f(G)-f\left(G^{\prime}\right)\right\|_1 } .
\end{equation}
\end{defn}

\begin{lem}
(Post-Processing Immunity\cite{TCS-042}) Let $\mathcal{F}:\mathcal{S} \rightarrow \mathcal{Y}$ be a randomized mapping and $\mathcal{M}$ satisfy $\varepsilon$-DP. For the mechanism $\mathcal{F} \circ \mathcal{M}$, it also satisfies $\varepsilon$-DP.
\end{lem}
Post-processing immunity is an inherent feature of any DP mechanism. It maintains the established privacy guarantees, even when arbitrary analysis is performed on the results of private queries.

\begin{lem}
\label{lem:comp}
(Sequential Composition\cite{TCS-042}) Let $\mathcal{M} = \{\mathcal{M}_1,...,\mathcal{M}_k\}$ where $\mathcal{M}_i$ satisfies ${\varepsilon}_i$-DP. $\mathcal{M}$ satisfies $\sum_{i=1}^{k} {\varepsilon}_i$-DP, if every $\mathcal{M}_i$ is performed on the same database.
\end{lem}

\subsection{Local Differential Privacy}
In local differential privacy, there is no trustworthy curator when executing a query on a database. In this model, each record is maintained independently by its holder. Without additional trust expenditure, every data holder transmits a perturbed version of their data to an untrusted curator for analyzing. For a distributed graph $G$, each vertex independently manages its neighbor list. We use $\mathbf{N}_u \in \{0,1\}^n$ to denote the view (its neighbors) of the vertex $u$. For example, for $i \in [n]$, if $\mathbf{N}_u{[i]}=1$ or $0$, indicates that the edge $e(u,i)$ exists or not, respectively. Then, the adjacency matrix $\mathbf{A}$ of $G$ can be denoted as $\{\mathbf{N}_1,...,\mathbf{N}_n\}$. We can define LDP under distributed graph. 
\begin{defn}
(Local Edge Differential Privacy) Suppose $G$ and $G^\prime$ differ in one edge. Let $\mathbf{N}_u, \mathbf{N}_u^\prime \in \mathcal{X}^{n}$ be two neighbor lists of any vertex $u$ in $G, G^\prime$, respectively. We can say $\mathcal{M}: \mathcal{X}^{n} \rightarrow \mathcal{X}^{n}$ satisfies $\varepsilon$-LDP, if for any possible $S \in range(\mathcal{M})$, have
\begin{equation}
        \operatorname{Pr}[\mathcal{M}\left(\mathbf{N}_u\right) \in S] \leq \exp (\varepsilon) \times \operatorname{Pr}\left[\mathcal{M}\left(\mathbf{N}_u^{\prime}\right) \in S\right].
\end{equation}
\end{defn}
In addition to Laplace noise, the randomized response \cite{warner1965randomized} can also be used to provide the LDP guarantee. Let $\mathcal{R}$ be a randomized response mechanism, for a binary value $b_i$
\begin{equation}
    \mathcal{R}(b_i)= \begin{cases}b_i & \text { with probability } 1-p \\ 1-b_i & \text { with probability } p\end{cases}.
\end{equation}
If $p=\frac{1}{1 + \exp (\varepsilon)}$, $\mathcal{M}=\{\mathcal{R}_1,...,\mathcal{R}_n\}$ satisfies $\varepsilon$-LDP.

With randomized response, LDP can provide $\varepsilon$-LDP for all vertices if their neighbor lists are independent. However, this independence may not be established since two vertices share one edge. Specifically, if $\mathbf{N}_u{[i]}=1$ or $0$, there must be $\mathbf{N}_i{[u]}=1$ or $0$, respectively. The privacy guarantee for a specific edge is influenced by two vertices, which increases the associated privacy risk. Previous studies have suggested that each vertex can report a subset of its neighbors to preserve independence \cite{qin2017generating, imola2021locally,eden2023triangle}. For instance, given an undirected graph, its adjacency matrix $\mathbf{A}$ is symmetric. For vertex $u \in [n]$, it only reports neighbors $\mathbf{N}_u{[u+1:]}$ with $\mathcal{R}$. The $\varepsilon$-LDP privacy guarantee can be preserved for each edge.

Intuitively, if each vertex chooses to report all its neighbors, and with $\mathcal{R}$ applied to each edge, the $2\varepsilon$-LDP can be held for each edge. 

\section{Graph Aggregation Approach} \label{sec:gam}
In this section, we introduce a graph aggregation approach designed to support distance queries on synthetic graphs. This approach addresses a challenge identified by Qin et al. \cite{qin2017generating}, specifically that direct application of randomized responses tends to produce a denser graph than the original. To mitigate this effect, our approach utilizes bitwise operations to refine the graph structure.

\subsection{The Motivation}
Let us review the issue highlighted by Qin et al. \cite{qin2017generating}. Each vertex only owns a limited view (its neighbors) of the whole graph. An untrusted curator, tasked with aggregating this local information to create a synthetic graph, requests each vertex to submit their neighbor list. With privacy concerns, each vertex employs a randomized response technique to perturb its entire list of neighbors. For each neighbor relationship of $u$
\begin{equation}
    \mathcal{R}(\mathbf{N}_u{[i]})= \begin{cases}\mathbf{N}_u{[i]} & \text { with probability } 1-p \\ 1-\mathbf{N}_u{[i]} & \text { with probability } p\end{cases}.
\end{equation}

By aggregating all neighbor lists from all local vertices, the curator can obtain a masked graph. However, the expected density of this graph is $\gamma(1-p)+(1-\gamma)p$ if the original density is $\gamma$. For example, if $\gamma = 0.1$ and $p=0.1$, the expected density is $0.18$, which is approximately twice the original value. These extra edges will severely destroy the graph structure, compared with the local graph. With this naive method, the density remains constant only under conditions where $p=0$ or $\gamma=1/2$. However, $p=0$ means serious privacy leakage, and $\gamma=1/2$ is too strict for real-world graphs.

\subsection{Graph Aggregation}
Algorithm \ref{alg:1} outlines our two-round graph aggregation process. In the first round (Steps $1$ to $6$), each local vertex sends their degree, adjusted with Laplace noise $\frac{2}{\varepsilon_1}\times \text{Lap}(1)$, to the third curator. The curator aggregates these to estimate the density as $\hat{\gamma} \leftarrow \frac{\sum_{i=1}^n\hat{d_i}}{n(n-1)}$ and then broadcasts this estimate to each vertex. 

In the second round (Steps $7$ to $17$), each vertex perturbs all their neighbor relationships through a randomized response $\mathcal{R}$. After randomization, each vertex transmits their noisy neighbor list $\hat{\mathbf{N}}_i$ to the curator. Since each relationship is reported twice, the curator confirms the existence of an edge using the $\textbf{AND}$ operator. That is, an edge $e(u,v)$ exists if both corresponding vertices $u$ and $v$ report it (i.e., $\hat{\mathbf{N}}_u[v]$ and $\hat{\mathbf{N}}_v[u]$ both equal to 1).  In this case, the curator records the edge's existence as $\hat{\mathbf{A}}{[u][v]} = 1$.

\begin{algorithm}            
  \caption{Graph Aggregation}
  \label{alg:1}
  \begin{algorithmic}[1]
  \renewcommand{\algorithmicrequire}{\textbf{Input:}}
  \renewcommand{\algorithmicensure}{\textbf{Output:}}
  \REQUIRE Privacy parameters $\varepsilon_1\text{ and } \varepsilon_2$\tikzmark{right1}
  \ENSURE  Noisy graph $\hat{\mathbf{A}}$ 
  
  \tikzmark{top1}
    \FOR{\text{each vertex} $i=1$ \textbf{to} $n$}

        \STATE $\hat{d_i}\leftarrow d_i+\frac{2}{\varepsilon_1}\times \text{Lap}(1)$ \tikzmark{right}
        \STATE send $\hat{d_i}$ to curator \tikzmark{right3}
    \ENDFOR\tikzmark{bottom1}
    \STATE $\hat{\gamma} \leftarrow \frac{\sum_{i=1}^n\hat{d_i}}{n(n-1)}$ \tikzmark{top3}
    \STATE send $\hat{\gamma}$ to vertices \tikzmark{bottom3}
    \STATE $p \leftarrow \frac{1}{\exp{(\varepsilon_2)}+1}$ \tikzmark{top2}
    \FOR{$i=1$ \textbf{to} $n$}
        \STATE $\hat{\mathbf{N}}_i \leftarrow \left(\mathcal{R}\left(\mathbf{N}_i{[1]}\right),...,\mathcal{R}\left(\mathbf{N}_i{[n]}\right)\right)$\tikzmark{right2}
        \STATE send $\hat{\mathbf{N}}_i$ to curator
    \ENDFOR\tikzmark{bottom2}
    \STATE \tikzmark{top4}
    \FOR{$i=1$ \textbf{to} $n$} 
        \FOR{$j=i+1$ \textbf{to} $n$} 
            \STATE $\hat{\mathbf{A}}{[i][j]} \leftarrow \hat{\mathbf{N}}_i{[j]}\bigcap \hat{\mathbf{N}}_j{[i]}$ \tikzmark{right4}
            \STATE $\hat{\mathbf{A}}{[j][i]} \leftarrow \hat{\mathbf{N}}_i{[j]}\bigcap \hat{\mathbf{N}}_j{[i]}$
        \ENDFOR
    \ENDFOR 
    \STATE \tikzmark{bottom4}
  \RETURN $\hat{\mathbf{A}}$
  \end{algorithmic}
  \AddNote{top1}{bottom1}{right1}{by vertices.}
  \AddNote{top2}{bottom2}{right2}{by vertices.}
  \AddNote{top3}{bottom3}{right3}{by curator.}
  \AddNote{top4}{bottom4}{right4}{by curator.}
\end{algorithm}

\begin{figure}
    \centering
    \includegraphics[width=0.9\linewidth]{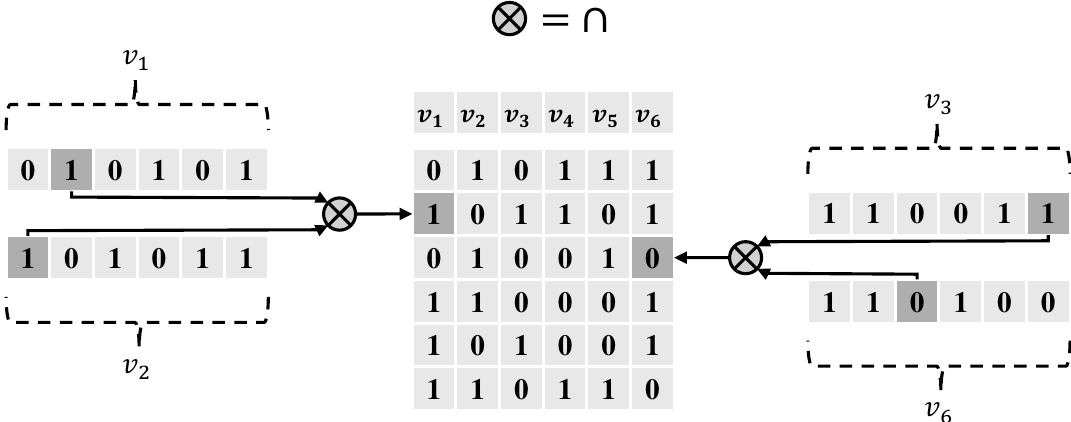}
    \caption{An example of using $\textbf{AND}$ operation to tune the edges}
    \label{fig:cap}
\end{figure}

\begin{exmp}

Figure \ref{fig:cap} illustrates the use of the $\textbf{AND}$ operation to tune the existence of edges. In the second round, vertices $v_1, v_2, v_3,$ and $v_4$ report their noisy neighbor lists. Using the $\textbf{AND}$ operation, the curator concludes that the edge $e(v_1, v_2)$ exists, as $\hat{\mathbf{N}}_1[2] \cap \hat{\mathbf{N}}_2[1] = 1$. Conversely, the edge $e(v_3, v_6)$ is determined to not exist, since $\hat{\mathbf{N}}_3[6] \cap \hat{\mathbf{N}}_6[3] = 0$.
\end{exmp}

\begin{thm}
Algorithm \ref{alg:1} satisfies $2(\varepsilon_1 + \varepsilon_2)$-LDP for every edge.
\end{thm}
$Proof$: the first round provides $2\varepsilon_1$-LDP for edges, since two vertices share each edge. In the second round, each edge is reported with $\mathcal{R}$ twice. Then, the $\textbf{AND}$ operation executed by the curator qualifies as post-processing. With Lemma \ref{lem:comp}, Algorithm \ref{alg:1} satisfies $2(\varepsilon_1 + \varepsilon_2)$-LDP for all edges.
$\hfill\blacksquare$

\begin{thm}\label{lem:err1}
When $\varepsilon_2 = \ln{\left(\frac{1}{2\hat{\gamma}}-1 \right)}$, the expectation of the density error $\mathbf{E}{[\Bar{\gamma}-\gamma]}=O\left(\frac{1}{\varepsilon_1^2 n(n-1)^2}\right)$, where $\bar{\gamma}$ is the density of graph $\hat{\mathbf{A}}$.
\end{thm}

$Proof$: with the $\textbf{AND}$ operation, we have
\begin{align}
    \mathbf{P}\left(\hat{\mathbf{A}}{[i][j]} =1\mid \mathbf{N}_i{[j]}=1\right) &= \left(1-p \right)^2,\\
    \mathbf{P}\left(\hat{\mathbf{A}}{[i][j]} =1\mid \mathbf{N}_i{[j]}=0\right) &= p^2.
\end{align}
Thus, we have the expected $\bar{\gamma}$
\begin{align}
    \bar{\gamma} &= \left(1-p\right)^2\cdot\gamma + \left(1-\gamma\right)\cdot p^2,\\
    \mathbf{E}\left[\bar{\gamma}\right] &= \mathbf{E}\left[\gamma + p^2-2p\gamma \right] \\
            &= \mathbf{E}\left[\gamma\right] + \mathbf{E}\left[p^2\right] - \mathbf{E}\left[2p\gamma \right].
\end{align}
To make sure $\bar{\gamma}$ is similar to $\gamma$, we assume $p=a\hat{\gamma} + b$ where $a$ and $b$ are constants. Since 
\begin{equation}
    \hat{\gamma} = \frac{\sum_{i=1}^{n}d_i}{n(n-1)} + \frac{2\sum_{i=1}^{n}X_i}{\varepsilon_1 n(n-1)} 
                =\gamma + \frac{2\sum_{i=1}^{n}X_i}{\varepsilon_1 n(n-1)} ,
\end{equation}
where $X_i$ is a Laplace random variable with mean $0$ and variance $2$. Thus, we have 
\begin{align}
    \text{var}\left[p\right] &= a^2 \cdot \text{var}\left[\hat{\gamma}\right]
            = \frac{4a^2}{\varepsilon_1^2 n(n-1)^2},\\
    \mathbf{E}\left[p^2\right] &=\text{var}\left[p\right] + \left(\mathbf{E}\left[p\right]\right)^2\\
            &= \frac{4a^2}{\varepsilon_1^2 n(n-1)^2} + \left(a\gamma + b\right)^2.
\end{align}
With above results, we can have
\begin{equation}
    \mathbf{E}\left[\bar{\gamma}\right] = \gamma + \frac{4a^2}{\varepsilon_1^2 n(n-1)^2} + \left(a\gamma + b\right)^2-2\gamma\left(a\gamma + b\right).
\end{equation}
For $a=2,b=0$, we can get small error
\begin{equation}
    \mathbf{E}\left[\bar{\gamma}-\gamma \right] = \frac{16}{\varepsilon_1^2 n(n-1)^2}.
\end{equation}
At this moment, $p=2\hat{\gamma}$, and thus $\varepsilon_2 = \ln{\left(\frac{1}{2\hat{\gamma}}-1 \right)}$ holds.
$\hfill\blacksquare $

\begin{algorithm}
  \caption{Improved Graph Aggregation (Partial)}
  \label{alg:2}
  \begin{algorithmic}[1]
  \renewcommand{\algorithmicrequire}{\textbf{Input:}}
  \renewcommand{\algorithmicensure}{\textbf{Output:}}
  \REQUIRE Privacy parameters $\varepsilon_1\text{ and } \varepsilon_2$
  \ENSURE  Noisy graph $\hat{\mathbf{A}}$
    \STATE $\alpha = \frac{2\hat{\gamma} + p -2}{2p-2}$ \tikzmark{top1}
    \FOR{$i=1$ \textbf{to} $n$}
        \FOR{$j=i+1$ \textbf{to} $n$}
            \STATE $b \leftarrow \text{Bernoulli}(\alpha)$
            \IF{ $b=1$}
                \STATE $\hat{\mathbf{A}}{[i][j]} \leftarrow \hat{\mathbf{N}}_i{[j]}\bigcap \hat{\mathbf{N}}_j{[i]}$\tikzmark{right1}
                
            \ELSE
                \STATE $\hat{\mathbf{A}}{[i][j]} \leftarrow \hat{\mathbf{N}}_i{[j]}\bigcup \hat{\mathbf{N}}_j{[i]}$
                
            \ENDIF
        \ENDFOR
    \ENDFOR \tikzmark{bottom1}
  \RETURN $\hat{\mathbf{A}}$
  \end{algorithmic}
  \AddNote{top1}{bottom1}{right1}{by curator.}
\end{algorithm}

When $\varepsilon_2$ is fixed, setting $p = 2\hat{\gamma}$ allows us to achieve $\hat{\mathbf{A}}$ with a minimal density error. However, under these conditions, the privacy level also depends on the local graph density $\gamma$. Graphs with low density offer weak privacy guarantees, while dense graphs provide strong protection. It is important to note that for graphs where $\gamma > 1/2$, we can opt to aggregate their complement graphs instead.

To reduce the strong dependency of $\varepsilon_2$ on $\hat{\gamma}$, we have enhanced our aggregation method with Algorithm \ref{alg:2}. This algorithm details the curator's actions in the second round. Since the other steps are identical to those in Algorithm \ref{alg:1}, they are omitted for simplicity. The key modification in Algorithm \ref{alg:2} involves the introduction of a Bernoulli random variable $b$, which determines whether to use $\textbf{AND}$ or $\textbf{OR}$. The curator aggregates $\hat{\mathbf{N}}_i{[j]}$ and $\hat{\mathbf{N}}_j{[i]}$ using $\textbf{AND}$ with probability $\alpha$, and $\textbf{OR}$ with probability $1-\alpha$.

\begin{thm}

With $\alpha = \frac{2\hat{\gamma} + p -2}{2p-2}$ and $\varepsilon_2 \geq \ln{\left(\frac{1}{2\hat{\gamma}}-1 \right)}$ if $\gamma \leq \frac{1}{2}$ or $\varepsilon_2 \geq \ln{\left(\frac{1}{2\left(1-\hat{\gamma}\right)}-1 \right)}$ if $\frac{1}{2} \leq \hat{\gamma}\leq 1$, the expectation of the density error $\mathbf{E}{[\Bar{\gamma}-\gamma]}=O\left(\frac{1}{\varepsilon_1^2 n(n-1)^2}\right)$. 
\end{thm}
$Proof$: similar to Theorem \ref{lem:err1}, we have the following facts:
\begin{align}
    \mathbf{P}\left(\hat{\mathbf{A}}{[i][j]} =1\mid \mathbf{N}_i{[j]}=1\right) &= 2\alpha p^2 -2\alpha p +1 -p^2,\\
    \mathbf{P}\left(\hat{\mathbf{A}}{[i][j]} =1\mid \mathbf{N}_i{[j]}=0\right) &= 2\alpha p^2 -2\alpha p -p^2 +2p.
\end{align}
Let $\alpha = \frac{2\hat{\gamma} + p -2}{2p-2}$, we have
\begin{align}
    \bar{\gamma} &= \gamma + 2\alpha p^2 -2\alpha p - p^2 +2p-2\gamma p\\
                &= 2p\left( \hat{\gamma}-\gamma \right).
\end{align}
Since $p$ should be related to $\gamma$, we assume that $p=a\hat{\gamma} + b$. Similar to the proof of Theorem \ref{lem:err1}, we have  
\begin{equation}
    \mathbf{E}\left[\bar{\gamma}-\gamma \right] = \frac{8a}{\varepsilon_1^2 n(n-1)^2}.
\end{equation}
With $0\leq \alpha \leq 1$, we can obtain $p \leq 2\hat{\gamma}$ if $\hat{\gamma}\leq \frac{1}{2}$, or $p \leq 2\left(1-\hat{\gamma}\right)$ if $\frac{1}{2} \leq \hat{\gamma}\leq 1$. 

Thus, for $p=\frac{1}{\exp{\left(\varepsilon_2\right)}-1}$, we have
\begin{equation}
    \varepsilon_2 \geq \ln{\left(\frac{1}{2\hat{\gamma}}-1 \right)},
\end{equation}
  if $\hat{\gamma}\leq \frac{1}{2}$, or 
\begin{equation}
    \varepsilon_2 \geq \ln{\left(\frac{1}{2\left(1-\hat{\gamma}\right)}-1 \right)},
\end{equation}
  if $\frac{1}{2} \leq \hat{\gamma}\leq 1$.
$\hfill\blacksquare $
\begin{rem}
Our graph aggregation approach is specifically designed for undirected graphs. If applied to a directed graph $G$, this approach would remove the directed nature of its edges, transforming $G$ into an undirected graph.
\end{rem}
\subsection{Limitation on Distance Query}
The main limitation of the above method, shared by many synthetic graph generation approaches, is the inherent noise introduced during the initial aggregation of neighbor lists. Methods such as RNL \cite{qin2017generating} and probability-based graph generation \cite{seshadhri2012community} introduce randomness when collecting or generating edges, which can unpredictably disrupt community structures. These structures are critical, as shortest paths between vertices often depend on them. Distance queries are highly sensitive to changes in graph structure, and this sensitivity amplifies the impact of noise originating from the graph generation process. As a result, it becomes difficult to explain or improve the expected utility of distance query results.

Instead of generating synthetic graphs for querying, some studies focus on specific tasks, such as clustering coefficient estimation \cite{ye2022lfgdpr}, subgraph counting (especially for triangle) \cite{imola2022communicationefficient}, perform subqueries that are then aggregated to achieve an overall estimation. Following a similar routine, aggregating the local distance subquery results from each vertex is a sensible strategy. This method helps avoid the unpredictability associated with local and global community structures in graphs.

\section{Neighbor Aggregation Approach} \label{sec:nam}
In this section, we introduce a neighbor aggregation method designed to address distance queries while providing a level of interpretability. 
\subsection{The Intuition}
Let $f$ be a distance query. For vertices $s$ and $t$, their distance $f(s,t)$ can be aggregated by subqueries
\begin{equation}
    f(s,t) \leq f(s,u) + f(u,t),
\end{equation}
where $u$ is an intermediate vertex. If $u$ is located on the shortest path between $s$ and $t$, the equation is satisfied. By strategically selecting an appropriate $u$, we can accurately aggregate the results to determine the distance.

Note that we hold an assumption in this paper that third-party adversaries cannot track traffic to reveal relationships. Thus, we can focus on the privacy risk from the query results. And this assumption is feasible through the use of a virtual private network in real-world. 

From vertex $s$'s perspective, communication is limited to its neighbors. If $s$ wants to know the distance to $t$ (where $t$ is not a neighbor of $s$), the most effective method is to inquire about the distances from its neighbors to $t$. For any neighbor $u$, the relationship $f(s,t)\leq 1+f(u,t)$ holds. Let us use $N(\cdot)$ to denote the neighbor set. Therefore, $s$ can aggregate all answers from its neighbors $N(s)$
\begin{equation}
    f(s,t) = 1+ \min_{u\in N(s)}\{f(u,t)\}.
\end{equation}

Neighbors can recursively perform the process to obtain the final result. However, two potential problems complicate this approach.
\begin{itemize}[leftmargin=0.3cm]
    \item $\textbf{Infinite-loop trap.}$ Cycles are common in graphs, resulting in distance queries being cyclically forwarded among vertices, thereby making it difficult to terminate the process.
    \item $\textbf{Uncertainty in waiting for answers.}$ When a vertex receives a distance query, it must wait for responses from all its neighbors, except the one that sent the query to this vertex. This recursive process can significantly increase the time overhead, especially in deeply structured graphs. Given each vertex's limited perspective, the waiting time is unpredictable. Furthermore, if messages are lost in transit, a vertex may end up waiting indefinitely.

\end{itemize}

\subsection{Neighbor Aggregation} \label{subsec:na}

\begin{algorithm}
  \caption{Neighbor Aggregation}
  \label{alg:na}
  \begin{algorithmic}[1]
  \renewcommand{\algorithmicrequire}{\textbf{Input:}}
  \renewcommand{\algorithmicensure}{\textbf{Output:}}
  \REQUIRE Initial noisy distance $\left\{\hat{\mathbf{D}}_1^{\left(0\right)},...,\hat{\mathbf{D}}_n^{\left(0\right)}\right\}$, distance threshold $T$
  \ENSURE  All-pair distances $\left\{\hat{\mathbf{D}}_1^{\left(T-1\right)},...,\hat{\mathbf{D}}_n^{\left(T-1\right)}\right\}$

    \FOR{$k=1$ \textbf{to} $T-1$}
        \FOR{\text{each vertex} $i=1$ \textbf{to} $n$}
            \STATE $\hat{\mathbf{D}}_i^{\left(k\right)} \leftarrow \text{Agg}\left(\left\{\hat{\mathbf{D}}_1^{\left(k-1\right)},...,\hat{\mathbf{D}}_n^{\left(k-1\right)}\right\}\right)$
        \ENDFOR
    \ENDFOR
  \RETURN $\left\{\hat{\mathbf{D}}_1^{\left(T-1\right)},...,\hat{\mathbf{D}}_n^{\left(T-1\right)}\right\}$
  \end{algorithmic}
\end{algorithm}

Algorithm \ref{alg:na} outlines the neighbor aggregation method, where the central idea is to continuously update global information by aggregating local data from neighbors, analogous to the learning mechanism in GNN \cite{sajadmanesh2023gap}. The key components of this method include the initial local distance design for each vertex and the implementation of the $\text{Agg}(\cdot)$ function.

$\textbf{Initialize Local Distance.}$ For each vertex in a local graph, they only know their neighbors. From their perspective, the distances to the other vertices are uncertain, except the distances to the neighbors, which are $1$. Let $\mathbf{D}_u^{\left(0\right)}$ denote the initial distances from the vertex $u$ to all other vertices. For any neighbor $j \in N(u)$, $\mathbf{D}_u^{\left(0\right)}{[u]} = 1$; for non-neighbors, $\mathbf{D}_u^{\left(0\right)}{[u]} = \infty$. And for $u$ itself, $\mathbf{D}_u^{\left(0\right)}{[u]} = 0$.

Subsequently, each vertex sends the distances to neighbors for aggregation. Since our privacy-preserving target is each edge in the graph, vertices must conceal their initial distance vector to prevent privacy leakage. Without such concealment,  directly exposing distances of '1' could inadvertently reveal neighbor relationships. To perturb these vectors effectively, two common methods are employed: Laplace and Randomized Response (RR).
\begin{itemize}[leftmargin=0.3cm]
  \item $\textbf{Laplace.}$ The vector $\mathbf{X}$ is an $n$-dimensional random variable, with each variable being independent and following a Laplace distribution characterized by a mean of $0$ and a scale parameter of $1$. We can conceal edges by 
  \begin{equation}
      \hat{\mathbf{D}}_u^{\left(0\right)} \leftarrow \mathbf{D}_u^{\left(0\right)} + \frac{T-1}{\varepsilon} \cdot \mathbf{X},
  \end{equation}
  where $\varepsilon$ is the privacy parameter, and $T$ is the distance threshold that replaces $\infty$ to avoid unbounded sensitivity.  
  
  \item $\textbf{RR.}$ Let $\mathcal{R}$ be a RR mechanism and $Unif(\cdot)$ be a uniform sampling function. We can conceal edges by giving $p=\frac{T}{e^{\varepsilon}+T-1}$
  
  \begin{equation}
      \mathcal{R}\left(x\right)=\left\{\begin{array}{ll}
                x & \text { with probability } 1-p \\
                Unif\left(\left[T\right]\right) & \text { with probability } p
                \end{array} ,\right.
  \end{equation}
    \begin{equation}
        \hat{\mathbf{D}}_u^{\left(0\right)} \leftarrow \left[ \mathcal{R}\left(\mathbf{D}_u^{\left(0\right)}{[1]} \right),...,\mathcal{R}\left(\mathbf{D}_u^{\left(0\right)}{[n]} \right)   \right].
    \end{equation}
\end{itemize}
We will discuss the differences between the two methods later in this subsection \ref{subsec:ua}.

$\textbf{Aggregation Process.}$ At first, each vertex just knows the exact distances to their neighbors; distances to all other vertices remain unknown. To manage these unknown distances, we use $T$ as a constraint. To update their unknown distances, all vertices can aggregate the distance vectors from their neighbors. The process can be described by the following formula.
\begin{equation}
\hat{\mathbf{D}}_u^{\left(k\right)}{\left[j\right]} \leftarrow \min\left\{ \min_{i\in N(u)}\left\{\hat{\mathbf{D}}_i^{\left(k-1\right)}{\left[j\right]}\right\}+1, \hat{\mathbf{D}}_u^{\left(k-1\right)}{\left[j\right]}\right\},
\end{equation}
where $j \in [n]\textbackslash N(u)$ and $k \geq 1$ denotes the $k$-th aggregation.

In the $k$-th aggregation, each vertex $u$ receives $|N(u)|$ messages containing distance vectors from its neighbors. These vectors are the latest knowledge that each neighbor has regarding their distances to all other vertices. Therefore, the updated distance from $u$ to another vertex $j$ should be calculated as the minimum of the sum of $u$'s distance to its neighbors and their distances to $j$. $\hat{\mathbf{D}}_u^{\left(k-1\right)}$ contains the results from the previous round of aggregation, ensuring that the vertex $u$ progressively acquires the optimal global distances to all vertices throughout the aggregation process.

Intuitively, this aggregation process resembles a breadth-first search. As each vertex continuously aggregates information from its neighbors, its perspective effectively expands outward by one hop with each aggregation. By the $k$-th aggregation, $u$ can obtain knowledge (initial distance vector) from its $k$-hop neighbors. This knowledge allows $u$ to identify its $(k+1)$-hop neighbors. Conversely, during the $k$-th aggregation, a vertex that is $k$-hop away from $u$ transmits its initial distance vector to $u$.

If no noise is injected, any vertex $u$ is able to find vertices at distance $k + 1$ at the $k$-th aggregation. Thus, each distance updated at an aggregation is optimal and will not change in later aggregations. However, with the introduction of random noise, the distance may be updated in each aggregation.

\begin{thm}\label{thm:ldp}
The Algorithm \ref{alg:na} satisfies $2\varepsilon$-LDP for each edge.
\end{thm}
$Proof$: the privacy guarantee is provided by the noisy initial distance vector $\left\{\hat{\mathbf{D}}_1^{\left(0\right)},...,\hat{\mathbf{D}}_n^{\left(0\right)}\right\}$. Each edge is disclosed twice by $\hat{\mathbf{D}}_i^{\left(0\right)}{[j]}$ and $\hat{\mathbf{D}}_j^{\left(0\right)}{[i]}$ where $i,j \in [n]$, it satisfies $2\varepsilon$-LDP. The later aggregation process is post-processing. Thus, this privacy guarantee holds.
$\hfill\blacksquare $

$\textbf{The trade-off for T.}$ In our setting, $T$ is the distance threshold that can limit our sensitivity to $T-1$. Large $T$ will introduce more rounds of aggregation, while small $T$ will cause excessive distance cropping, leading to larger errors. Diameter is an alternative option. We can predict the diameter by following diameter bound.

\begin{lem}\label{lem:dbound}
(Diameter bound \cite{mukwembi2012note}) Given $G$, a connected graph with $n$ vertices. For its diameter $d$
\begin{equation}
    d \leq \frac{3(n-t)}{\delta+1}+O(1),
\end{equation}
where $\delta$ is the minimum degree and $t$ represents the count of unique values in the degree sequence.
\end{lem}
Thus, the diameter can be estimated by an additional round of degree collection, similar to the first round in Algorithm \ref{alg:1}. If the privacy loss for degree collection is $\varepsilon_1$ and for neighbor aggregation is $\varepsilon_2$, the total privacy loss is $2(\varepsilon_1+\varepsilon_2)$. However, the bound given by Lemma \ref{lem:dbound} is not tight. Therefore, to facilitate experimental analysis, we empirically specify $T$ in the experiments through six degrees of separation \cite{elmacioglu2005six}.
\begin{rem}
In this paper, our privacy-preserving target is any neighbor relationship for every vertex. Before the aggregation process, each vertex $u$ has injected enough randomness into its neighbor relationships (the initial distance vector $\mathbf{D}_u^{\left(0\right)}$). Thus, each vertex cannot increase their confidence in their neighbors' privacy by receiving their neighbors' distance vectors. However, we cannot guarantee that the change in the distance vector will not expose indirect neighbor relationships ($2$-hop or more distant neighbors) in each aggregation. Fortunately, these indirect relationships, which are not the primary target of our privacy measures, generally hold less value than direct neighbor relationships.
\end{rem}

\subsection{Attack Resistance}\label{subsec:ar}
The perturbation used in the graph aggregation approach is similar to the neighbor aggregation approach. The key difference is that graph aggregation shares neighbor lists with all neighbors rather than just with a single curator. In fact, for each vertex, the initial distance vector $\mathbf{D}_u^{(0)}$ is equal to its neighbor list (just replace the value $T$ by $0$). Hence, both approaches comply with LDP.

In analyses involving multiple parties, such as Multi-Party Computation or federated learning, the assumption of participants being "honest but curious" is practical and common. Following this assumption, we consider all vertices to behave honestly during the execution but still attempt to extract private information from the aggregation process. However, such attempts fail due to the added noise in each local distance vector $\mathbf{D}_u^{(0)}$. We now discuss scenarios involving malicious vertices:

\textbf{Case 1: One Malicious Vertex.} Suppose vertex $m$ behaves maliciously with two possible goals: 1) extracting private edge information and 2) increasing errors in the aggregated results. Extracting private edge information is impossible since each vertex's neighbor relationships are concealed in the initial distance vectors $\mathbf{D}_v^{(0)}, v \in [n]$. Further aggregation does not disclose any additional neighbor relationship information. However, vertex $m$ can negatively affect the accuracy by initializing $\mathbf{D}_m^{(0)}$ incorrectly (e.g., with all zeros or ones). This manipulation can cause other vertices to underestimate distances, thus increasing overall error.

\textbf{Case 2: Multiple Malicious Vertices.} The worst-case scenario involves vertex $u$ having all neighbors except vertex $v$ being malicious. These malicious neighbors may try to determine if vertex $v$ is also a neighbor of vertex $u$. Such collusion attacks fail because all malicious vertices only receive the same perturbed distance vector $\hat{\mathbf{D}}_u^{(0)}$ from $u$, adding no extra privacy loss. However, if one malicious neighbor $t$ is also a neighbor of $v$, the privacy loss regarding the $u$-$v$ relationship is doubled. This scenario aligns with the upper bound privacy loss indicated by Theorem \ref{thm:ldp}. It is unnecessary to consider the scenario where all neighbors of $u$ are malicious, as collusion would straightforwardly reveal $u$'s neighbor list.

In conclusion, while malicious vertices may reduce the utility of all-pairs distance measurements, they do not compromise overall privacy.

\subsection{Utility Analysis}\label{subsec:ua}
In Subsection \ref{subsec:na} we provide two perturbation methods: Laplace and RR. These two have similar utility for some common queries in the local model, such as summation or counting. However, due to the complexity of the aggregation process, it is difficult to analyze the error bounds theoretically. Thus, we propose two random variables, $Y_1$ and $Y_2$, to support the empirical utility analysis. Section \ref{sec:exp} presents numerical simulations based on the following guarantees. We use $\min\{\cdot\}\_n$ to denote the minimum of $n$ independent random variables.

\begin{thm}\label{lem:lap}
For Laplace method, the distance $\hat{\mathbf{D}}_u^{\left(T-1\right)}{[j]} $, if $j$ is not the neighbor of $u$, can be described by random variable $Y_1=\min\left\{T, \min\left\{W+X\right\}_{n-2}\right\}$. Here, $W$ is a discrete random variable that depends on the graph structure around the vertex $u$, and $X$ is a Laplace random variable with mean $0$ and scale $\frac{T-1}{\varepsilon}$.
\end{thm}
$Proof$: Let $t$ be the true distance from vertex $u$ to vertex $j$. For $k$-th aggregation, $u$ can aggregate the neighbor relationships from $k$-hop neighbors. Without loss of generality, we assume $t< T-2$. 

\begin{itemize}[leftmargin=0.3cm]
    \item For $k < t-1$ or $k > t+1$,
    \begin{equation}
    \hat{\mathbf{D}}_u^{\left(k\right)}{\left[j\right]} \leftarrow \min\left\{ \hat{\mathbf{D}}_u^{\left(k-1\right)}{\left[j\right]}, \min\left\{X+T+k \right\}_{m_k}\right\},
    \end{equation}
    where $m_k$ is the number of $k$-hop neighbors of the vertex $u$.
    \item For $k =t-1$ or $k= t+1$, since there are some vertices that are neighbors of $j$, let $\omega = \min\left\{X+k+1 \right\}_{a_k}$,
    \begin{align}
     \hat{\mathbf{D}}_u^{\left(k\right)}{\left[j\right]} \leftarrow \min\left\{ \hat{\mathbf{D}}_u^{\left(k-1\right)}{\left[j\right]}, \min\left\{X+T+k \right\}_{m_{k}-a_k},\omega\right\},
    \end{align}
    where $a_{k}$ is the number of neighbors $j$ has, for which the distance to vertex $u$ is $k$.

    \item For $k=t$, there is an exception that the distance $\hat{\mathbf{D}}_j^{\left(0\right)}{\left[j\right]}$ will not propagate to $u$ since $j$'s neighbors will ignore it when aggregating. At this aggregation, let $\omega = \min\left\{X+k+1 \right\}_{a_k}$,
\begin{align}
 \hat{\mathbf{D}}_u^{\left(k\right)}{\left[j\right]} \leftarrow \min\left\{ \hat{\mathbf{D}}_u^{\left(k-1\right)}{\left[j\right]}, \min\left\{X+T+k \right\}_{m_{k}-a_k-1},\omega\right\}.
\end{align}
\end{itemize}

For clarity, we denote $a_{t-1},a_{t}$ and $a_{t+1}$ by $a_1, a_2$ and $a_3$, respectively. Let $ W $ be a random variable drawn from the distribution that represents the above constants. Specifically, we consider the histogram, which captures the frequency of occurrence of all constants (such as $ T+k $ or $ k+1 $), as a probability distribution. This histogram is detailed in Table \ref{tab:distri}. $\hat{\mathbf{D}}_u^{\left(T-1\right)}{[j]}$ can be represented by a random variable $Y_1$
\begin{equation}
    Y_1=\min\left\{T, \min\left\{W+X\right\}_{n-2}\right\},
\end{equation}
where the $n-2$ means that $u$ will aggregate distance vectors from all vertices except itself and $j$.
$\hfill\blacksquare $

\begin{thm}\label{lem:rr}
For RR method, the distance $\hat{\mathbf{D}}_u^{\left(T-1\right)}{[j]} $, is the value of $Y_2=\min\left\{T, \min\left\{W_1+X_1\right\}_{m},\min\left\{W_2+X_2\right\}_{a}\right\}$ if $j$ is not the neighbor of $u$. Here, $a+m=n-2$, $W_1$ and $W_2$ are two discrete random variables that depend on the graph structure around the vertex $u$, and $X_1$, $X_2$
\begin{align}
    P(X_1 = x) &= 
    \begin{cases}
        1 - \frac{T-2}{T-1}p & \text{if } x = 0, \\
        \frac{T-2}{T-1}p & \text{if } x = \text{Unif}([1-T, -1]),
    \end{cases} \\
    P(X_2 = x) &= 
    \begin{cases}
        1 - \frac{T-2}{T-1}p & \text{if } x = 0, \\
        \frac{T-2}{T-1}p & \text{if } x = \text{Unif}([1, T-1]).
    \end{cases}
\end{align}
  
\end{thm}

$Proof$: let us reformulate RR using random variables. For $\mathbf{D}_u^{\left(0\right)}{[j]}=T$, $\mathcal{R}\left(T\right)$ has the same distribution as $X_1+T$. And for $\mathbf{D}_u^{\left(0\right)}{[j]}=1$, $\mathcal{R}\left(1\right)$ has the same distribution as $X_2+1$. The distributions of $X_1$ and $X_2$ are listed above. Most of the proof is similar to Theorem \ref{lem:lap}, but using $X_1$ and $X_2$ replace $X$.
\begin{itemize}[leftmargin=0.3cm]
    \item For $k < t-1$ or $k > t+1$,
\begin{equation}
\hat{\mathbf{D}}_u^{\left(k\right)}{\left[j\right]} \leftarrow \min\left\{ \hat{\mathbf{D}}_u^{\left(k-1\right)}{\left[j\right]}, \min\left\{X_1+T+k \right\}_{m_k}\right\}.
\end{equation}

    \item For $k =t-1$ or $k= t+1$, $\hat{\mathbf{D}}_u^{\left(k\right)}{\left[j\right]}$ is the minimum of $\hat{\mathbf{D}}_u^{\left(k-1\right)}{\left[j\right]}$ and
\begin{align}
 \min\left\{ \min\left\{X_1+T+k \right\}_{m_{k}-a_k},\min\left\{X_2+k+1 \right\}_{a_k}\right\}.
\end{align}

    \item For $k=t$, $\hat{\mathbf{D}}_u^{\left(k\right)}{\left[j\right]}$ is the minimum of $\hat{\mathbf{D}}_u^{\left(k-1\right)}{\left[j\right]}$ and 
\begin{align}
 \min\left\{, \min\left\{X_1+T+k \right\}_{m_{k}-a_k-1},\min\left\{X_2+k+1 \right\}_{a_k}\right\}.
\end{align}
\end{itemize}

For clarity, we denote $a_{t-1},a_{t}$ and $a_{t+1}$ by $a_1, a_2$ and $a_3$, respectively. Let $a = \sum_{i=1}^{3}a_i$ and $m = n-2-a$, we have  
\begin{equation}
    Y_2=\min\left\{T, \min\left\{W_1+X_1\right\}_{m},\min\left\{W_2+X_2\right\}_{a}\right\},
\end{equation}
where $W = W_1\cup W_2$. 
$\hfill\blacksquare $

\begin{table}[h]
\centering
\caption{The histogram of all constants.}
\begin{tabularx}{\columnwidth}{Y|Y|Y|Y}
\toprule
\textbf{Constant} & $W$ & $W_1$ & $W_2$ \\
\midrule
$t$ & $a_1$ & $0$ & $a_1$ \\
$t+1$ & $a_2$ & $0$ & $a_2$ \\
$t+2$ & $a_3$ & $0$ & $a_3$ \\
\cdashline{1-4}
$T+1$ & $m_1$ & $m_1$ & $0$ \\
$T+2$ & $m_2$ & $m_2$ & $0$ \\
\cdashline{1-4} 
\multicolumn{4}{c}{$\cdots$} \\ 
\cdashline{1-4}
$T+t-1$ & $m_{t-1} - a_1$ & $m_{t-1} - a_1$ & $0$ \\
$T+t$ & $m_t - a_2 - 1$ & $m_t - a_2 - 1$ & $0$ \\
$T+t+1$ & $m_{t+1} - a_3$ & $m_{t+1} - a_3$ & $0$ \\
\cdashline{1-4} 
$T+t+2$ & $m_{t+2}$ & $m_{t+2}$ & $0$ \\
\multicolumn{4}{c}{$\cdots$} \\
$T+T-1$ & $m_{T-1}$ & $m_{T-1}$ & $0$ \\
\bottomrule
\end{tabularx}

\label{tab:distri}
\end{table}

\section{Experiments}\label{sec:exp}
In this section, we evaluate the utility of the Laplace and RR methods through numerical simulations. Subsequently, we compare the performance of our two approaches with LDPGen and RNL \cite{qin2017generating} in three real-world datasets. 

\subsection{Datasets and Metric}
We utilize three real-world datasets of varying sizes, described as follows:
\begin{itemize}[leftmargin=0.3cm]
    \item EIES \cite{freeman1979networkers}. This dataset represents a personal relationship graph among researchers. For our analysis, we ignore the weight information and utilize its complement graph, which comprises $34$ vertices and $87$ edges.
    \item Twitter \cite{fink2023centrality}. This is a directed Twitter interaction graph about the United States Congress. It contains $475$ vertices and $10,222$ edges.
    \item Facebook \cite{leskovec2012learning}. Originally an undirected social graph from the Facebook app, this dataset includes $10$ networks. We use the largest network that contains $1,034$ vertices and $26,749$ edges.
    
\end{itemize}

To evaluate the performance of our approaches on distance queries, we employ the Relative Absolute Mean Error (RAME) and the Mean Relative Error (MRE). RMAE captures the errors across all elements and is sensitive to fine-grained error distributions, whereas MRE is insensitive to individual element errors and reflects only the overall trend.
\begin{defn}
(RAME and MRE) Let $d_{u,v}$ be the distance from the vertex $u$ to $v$ and ${d}^\prime_{u,v}$ be the noisy distance. We have RAME $\eta_1$ and MRE $\eta_2$
\begin{align}
    \eta_1 &=\frac{1}{n^2 -n} \sum_{\substack{u, v \in G \\ u \neq v}} {\frac{\left|{d}^\prime_{u,v}-{d}_{u,v}\right|}{\left|{d}_{u,v}\right|}},
    \eta_2 = \frac{|\mathbf{\bar{d}}^\prime-\mathbf{\bar{d}}|}{|\mathbf{\bar{d}}|},
\end{align}
where $\mathbf{\bar{d}}$ and $\mathbf{\bar{d}}^\prime$ are the mean of all distances and all noisy distances, respectively.

\end{defn}

\begin{figure}[H]
    \centering
    \includegraphics[width=1\linewidth]{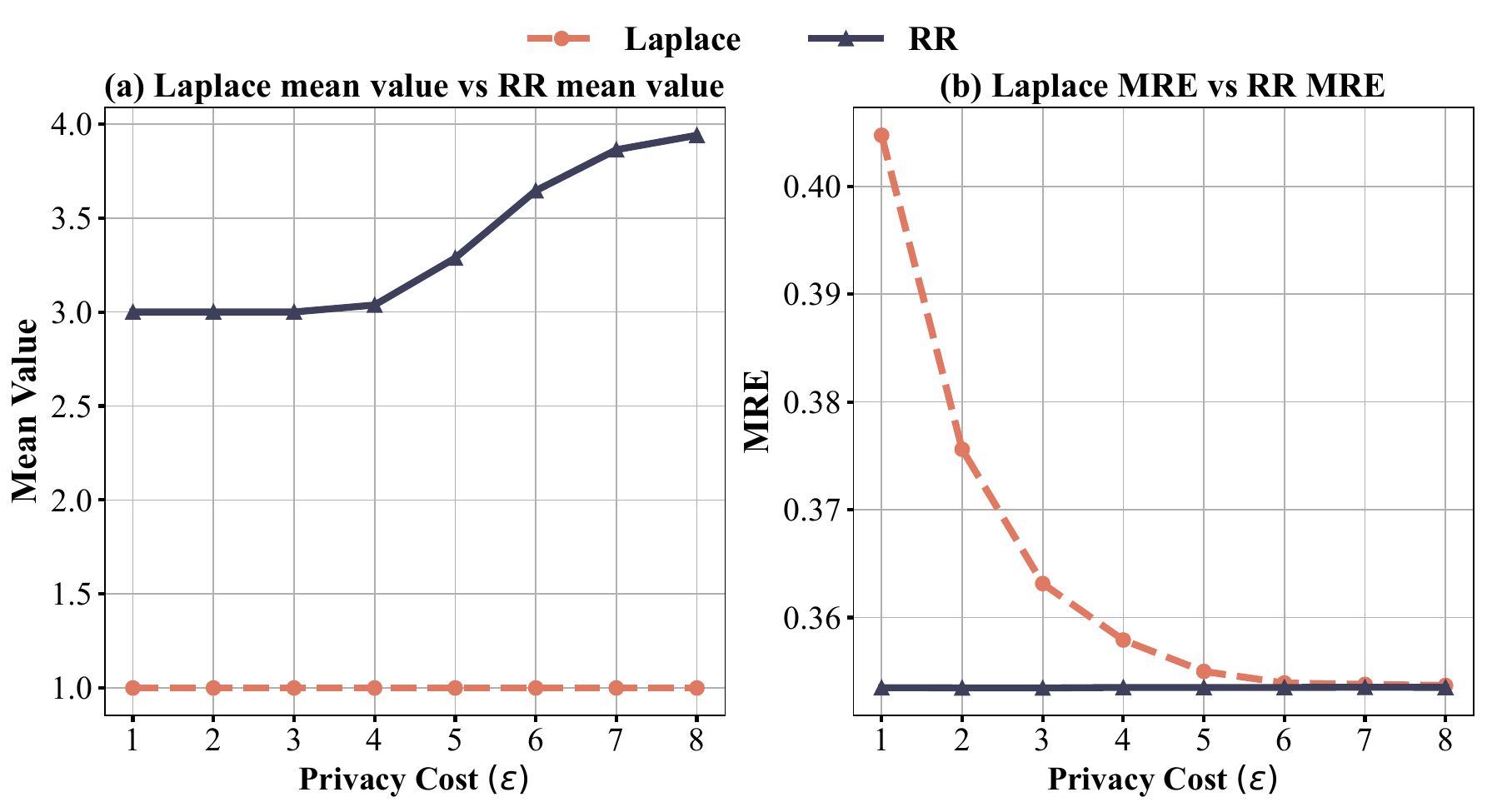}
    \caption{The simulation results for Laplace and RR in privacy cost $[1,8]$.}
    \label{fig:sim}
\end{figure}

\subsection{Simulation Results}

We conduct numerical simulations based on Theorems \ref{lem:lap} and \ref{lem:rr} to compare the utility of the Laplace and RR mechanisms. Given that the distributions of $X$, $X_1$, and $X_2$ are already defined, the primary focus shifts to determining the distributions of $W$, $W_1$, and $W_2$. As depicted in Table \ref{tab:distri}, these variables form histograms where $W=W_1 \cup W_2$. We approximate the histogram frequencies as probabilities and, for simplification, assume that they follow a uniform distribution.

We empirically set $T=6$ for our experiments, inspired by the concept of six degrees of separation. Assuming, without loss of generality, that the true distance between vertices $u$ and $v$ is $4$. To reduce randomness in our results, we set $n = 10,000$ and repeat the experiments $1,000$ times. Finally, we set $\varepsilon \in [1,8]$. This range of $\varepsilon$ does not represent a serious privacy risk but is chosen to better illustrate trends.

In Fig. \ref{fig:sim}(a), as $\varepsilon$ increases, the mean value of the RR method approaches $4$, reflecting an increase in accuracy. In contrast, the mean value of the Laplace method remains consistently close to $1$. This consistency occurs because the sample values from the Laplace method are predominantly negative, leading us to adjust for utility by setting any sample value less than $1$ equal to $1$. Furthermore, Fig. \ref{fig:sim}(b) illustrates the MRE of both the Laplace and RR methods when applied to the Facebook dataset. It is evident that the RR method consistently yields smaller errors compared to the Laplace method as $\varepsilon$ increases.

The underperformance of the Laplace method can be ascribed to its unbounded sample space. The $\min$ causes the expectation to be biased in a negative direction. Let $Y=\min\{X\}_n$, $X$ be an independent Laplace random variable with mean $0$ and variance $2b^2$. Using the properties of cumulative distribution functions and order statistics, we can get the expectation of $Y$
\begin{equation}
    \mathbf{E}[Y]=b\ln{\left(\frac{1}{2}-\frac{1}{n+1}\right)},
\end{equation}
if $n>1$.

We observe that $\mathbf{E}[Y] < 0$, a phenomenon that is likely to persist for $Y_1$ as well, despite the bias introduced by $W$ that tends to shift its expectation towards positive values. Consequently, we will employ the RR method in our subsequent experiments.
\begin{figure}
    \centering
    \includegraphics[width=1\linewidth]{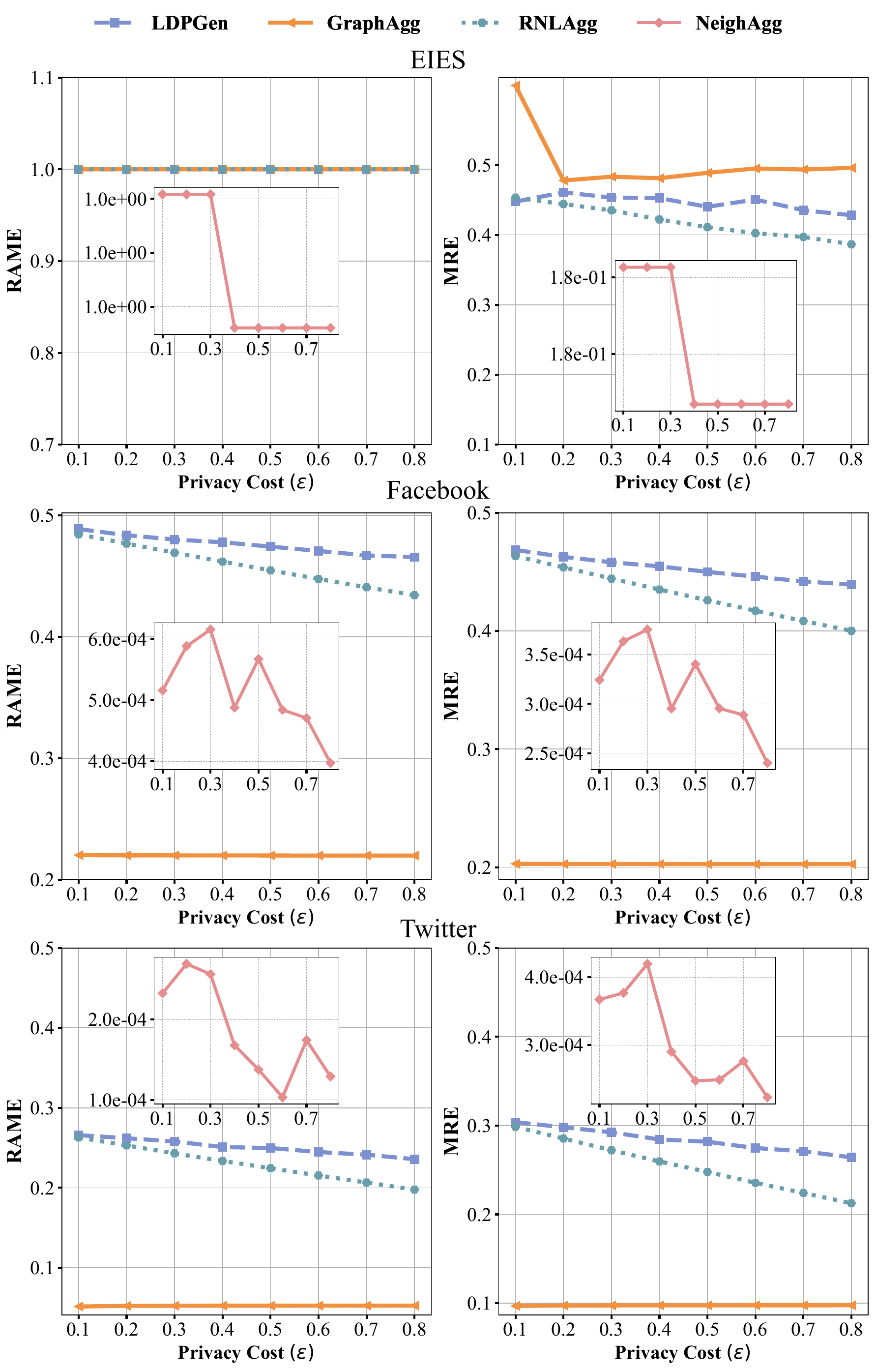}
    \caption{The RAME and MRE results for four approaches: LDPGen, GraphAgg, RNLAgg and NeighAgg under three real-world datasets: EIES, Facebook and Twitter when $\varepsilon$ changes from $0.1$ to $0.8$. The inset plot in each plot shows the trend of NeighAgg for clear visibility.}
    \label{fig:all_error}
\end{figure}

\subsection{Performance Results}

We perform our experiments on three real-world datasets using two our approaches: graph aggregation (GraphAgg) and neighbor aggregation (NeighAgg). Due to the lack of similar methods for distance queries, we select the classical LDPGen and RNLAgg \cite{qin2017generating} as our baselines for comparative analysis. The initial parameters for LDPGen are set according to its recommended settings: $K_0=2$, $\varepsilon_1= \varepsilon/2$, and $\varepsilon_2= \varepsilon/2$. 

For GraphAgg, we use Algorithm \ref{alg:1}. The privacy budget for the first round is $\varepsilon$. And for NeighAgg, with a fixed $T=6$, $\varepsilon/2$ directly represents the privacy budget utilized to initialize each disance vector. 
Since LDPGen, RNLAgg, and GraphAgg can generate synthetic graphs where certain vertex pairs become unreachable, we uniformly assign $T=6$ to represent the distance between such unreachable vertices.

Fig. \ref{fig:all_error} demonstrates the comparative performance of our two methodologies against established baselines. Across all graphs, there is a noticeable decreasing trend in errors as $\varepsilon$ increases. Notably, except for EIES, all methods adhere to the following pattern: NeighAgg $<$ GraphAgg $<$ RNLAgg $<$ LDPGen, with NeighAgg exhibiting significantly lower errors, approximately $10^{-4}$. It is important to mention that for GraphAgg, errors consistently reduce with higher values of $\varepsilon$. However, due to space constraints, we are unable to present these details as comprehensively as for NeighAgg. In the case of EIES, its atypical variations can be attributed to having fewer vertices and a higher susceptibility to noise interference. Nonetheless, NeighAgg maintains the lowest error rate among the compared methods in EIES.

\begin{figure}
    \centering
    \includegraphics[width=1\linewidth]{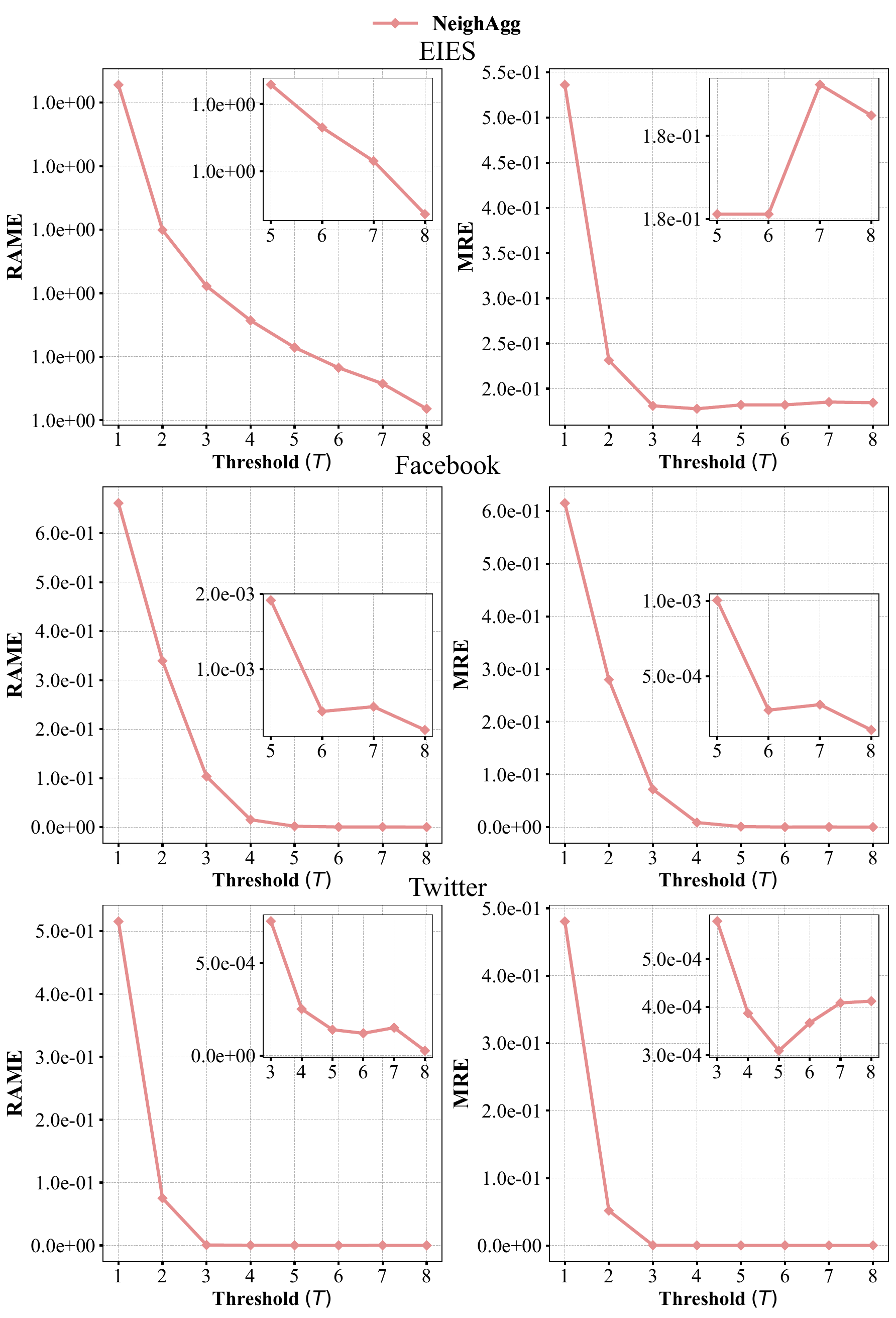}
    \caption{The RAME and MRE results for NeighAgg under three real-world datasets: EIES, Facebook and Twitter when  fixing $\varepsilon=0.1$ and $T$ changes from $1$ to $8$. The inset plot in each plot shows local trend of NeighAgg for clear visibility.}
    \label{fig:errorT}
\end{figure}

Fig. \ref{fig:errorT} illustrates the impact of varying $T$ values on the error. Across all plots, the error decreases as $T$ increases. However, a larger $T$ also results in higher computational costs, as more aggregation steps are required in NeighAgg, leading to a linear increase in runtime. Therefore, we identify $T=6$ as the optimal choice, striking a balance between low error and acceptable aggregation overhead.

\section{Related Work}\label{sec:RW}
\subsection{Synthetic Graph Generation}

Qin et al. \cite{qin2017generating} proposed a synthetic social graph generation method under LDP by combining the Randomized Neighbor List (RNL) and Degree-Based Graph Generation (DGG). They achieved a superior partition of the vertices using the $k$-means several times. Ju et al. \cite{ju2019generating} explored graph correlations and developed a correlation-based method for synthetic graph generation to address privacy leaks. Wei et al. \cite{wei2020asgldp} achieved preservation for graph statistics and attribute distributions by finely tuning the noise added. Focusing on the dynamic update problem of synthetic graphs, Hou et al. \cite{hou2023ppdu} proposed a new graph publishing mechanism that reduces noise dependency on graph scale through a privacy-preference-specifying mechanism, facilitating the release of dynamic graphs.

\subsection{Distance Query}

Current approaches that support differentially private distance queries are mostly based on the weight private graph assumption. This assumption can limit the global sensitivity to $1$ (or a unit). Using the graph assumption, Sealfon \cite{sealfon2016shortest} pioneered the release of all-pair distances, with the global sensitivity considered negligible. This work showed that the upper bound of the additional error is $O(n \log n / \varepsilon)$. Fan et al. \cite{fan2022distances} devised two strategies specifically for trees and grid graphs that limit errors to $O(\log^{1.5} n)$ and $\tilde{O}(n^{3/4})$, respectively. Furthermore, Fan et al. \cite{fan2022private} developed a method for privacy-sensitive distance queries via synthetic graphs, resulting in an error of $\tilde{O}(n^{1/2})$. With the same assumption, Chen et al. \cite{chen2023differentially} provided an algorithm for distance release, incurring an error $\tilde{O}(n^{2/3} / \varepsilon)$ and noted that the minimum error should be $\Omega(n^{1/6})$. They further reduced the error for bounded weights to roughly $n^{(\sqrt{17}-3)/2 + o(1)} / \varepsilon$. Leaving that assumption, Cai et al. \cite{cai2024shortest} examined scenarios involving more significantly varied neighboring weights, specifically where weights range from $[a, b]$ and differences can reach $b-a$. This variation introduces substantial global sensitivity. Consequently, their research focuses on mitigating the impact of noise on the shortest paths and distances.
\section{Conclusion}\label{sec:con}
In this paper, we propose two novel approaches for distance query under LDP. The first approach, graph aggregation, effectively addresses issues related to graph density by generating synthetic graphs through the innovative use of bitwise operations on RNL. Our second approach, neighbor aggregation, addresses the low utility typically observed in synthetic graph methods by enabling the continuous aggregation of distance vectors from each vertex’s neighbors. Both theoretical analysis and empirical evaluation confirm the robustness and effectiveness of the proposed approaches.

\section*{Acknowledgments}
This paper benefited from the use of OpenAI’s ChatGPT in polishing the language and improving clarity. No part of the technical contributions or experimental results was generated by AI models.

\bibliographystyle{IEEEtran}
\bibliography{references}{}

\end{document}